\documentclass[%
 reprint,
superscriptaddress,
 amsmath,amssymb,
 aps,prl,
]{revtex4-1}

\usepackage{todonotes}
\usepackage{graphicx}%
\usepackage{dcolumn}%
\usepackage{bm}%
\usepackage[normalem]{ulem}
\usepackage{microtype}
\usepackage{xcolor}
\usepackage{amsmath,amssymb,amsfonts,mathtools,cancel}
\usepackage{multirow}

\newcommand{\bc}[1]{{\color{black} #1 }}
\definecolor{DragonGreen}{RGB}{0,126,48}

\definecolor{Brownie}{RGB}{97,16,9}
 
\newcommand{\sm}{supplementary material~}

\usepackage[paperwidth=210mm,
            paperheight=297mm,
            left=20mm,
            top=10mm,
            textwidth=170mm,
            marginparsep=3mm,
            marginparwidth=30mm,
            textheight=730pt,
            footskip=50pt]
           {geometry}

\newcommand{\avg}[1]{\ensuremath{\langle{#1}\rangle}}
\newcommand{\abs}[1]{\ensuremath{\lvert {#1}\rvert}}

\newcommand{\gsl}{\ensuremath{\gamma_{\text{sl}}}}
\newcommand{\gp}{\ensuremath{\gamma_{\infty}}}
\newcommand{\ns}{\ensuremath{{n_{\text{s}}}}} %

\newcommand{\vs}{\ensuremath{v_{\text{s}}}}
\newcommand{\vl}{\ensuremath{v_{\text{l}}}}

\newcommand{\msl}{\ensuremath{\mu_{\text{sl}}}} %
\newcommand{\phis}{\ensuremath{\phi_{\text{s}}}} %
\newcommand{\phil}{\ensuremath{\phi_{\text{l}}}}

\newcommand{\tm}{\ensuremath{T_{\text{m}}}}

\newcommand{\kb}{\ensuremath{k_{\rm B}}}

\begin{document}

\preprint{APS/123-QED}

\title{Theoretical prediction of the homogeneous ice nucleation rate:\\
disentangling thermodynamics and kinetics}

\author{Bingqing Cheng}
\email{bingqing.cheng@epfl.ch}
 \affiliation{Laboratory of Computational Science and Modeling, Institute of Materials, {\'E}cole Polytechnique F{\'e}d{\'e}rale de Lausanne, 1015 Lausanne, Switzerland}%

\author{Christoph Dellago}
\affiliation{Faculty of Physics, University of Vienna, Boltzmanngasse 5, 1090 Vienna, Austria}

\author{Michele Ceriotti}
\affiliation{Laboratory of Computational Science and Modeling, Institute of Materials, {\'E}cole Polytechnique F{\'e}d{\'e}rale de Lausanne, 1015 Lausanne, Switzerland}%

\date{\today}%

\begin{abstract}
Estimating the homogeneous ice nucleation rate from undercooled liquid water is at the same time crucial for understanding many important physical phenomena and technological applications, and challenging for both experiments and theory.
From a theoretical point of view, difficulties arise due to the long time scales required, as well as the numerous nucleation pathways involved to form ice nuclei with different stacking disorders.
We computed the homogeneous ice nucleation rate at a physically relevant undercooling for a single-site water model,
taking into account the diffuse nature of ice-water interfaces,
stacking disorders in ice nuclei, and the addition rate of particles to the critical nucleus.
We disentangled and investigated the relative importance of all the terms, including interfacial free energy,
entropic contributions and the kinetic prefactor, that contribute to the overall nucleation rate. 
\bc{
There has been a long-standing discrepancy for the predicted homogeneous ice nucleation rates,
and our estimate is faster by 9 orders of magnitude compared with previous literature values.
Breaking down the problem into segments and considering each term carefully can help us understand where the discrepancy may come from and how to systematically improve the existing computational methods.}

\end{abstract}


\maketitle

\subsection{Introduction}

Nucleation of solids from liquids is commonplace in our daily lives,
and it has countless implications for science and technology~\cite{sosso2016crystal}:
Polymorph selection for pharmaceutical compounds~\cite{price2005crystalline},
the formation of amyloid plaques, which is related to Alzheimer's disease~\cite{hardy2002amyloid},
the growth of dendrites during solidification~\cite{flemings1974solidification}
are just a few examples.
The most obvious case is probably the 
nucleation of ice from undercooled liquid water.
It is not only an ubiquitous process in nature that influences global phenomena such as climate change, but also has many practical implications in refrigeration, anti-freezing, solidification and melting of solutions, as well as many other technological applications~\cite{oxtoby1992homogeneous,yi2012molecular,sosso2016crystal}.

Despite its pivotal importance, our
understanding of homogeneous ice nucleation and homogeneous nucleation in general is far from complete,
which is partly due to that fact 
the experimental investigation of dynamical nucleation processes is very difficult and often costly~\cite{pouget2009initial}. 
An alternative is to rely on atomistic simulations to study nucleation,
which has gained a lot of popularity in the last two decades~\cite{sosso2016crystal}.
However,
predicting ice nucleation rates using atomistic simulations is plagued by difficulties.
One major challenge arises from the inaccuracies in modelling the unique properties of water using either empirical potentials or \emph{ab initio} methods~\cite{jorgensen1983comparison,vega2009ice,vega2011simulating,paesani2009properties,reddy2016accuracy,morawietz2016van}.
On the other hand,
even when the same water model and the same thermodynamic conditions are assumed, nucleation rates predicted in different studies typically differ by as much as 5-10 orders of magnitude~\cite{sosso2016crystal}. 
This discrepancy is due to the fact that it is often necessary to evoke the standard form of classical nucleation
theory (CNT) in order to estimate quantities such as the nucleation barrier,
as the long time scale of nucleation rules out the option of brute force molecular dynamics simulations.
However, a number of approximations within CNT have been shown to be over-simplifications~\cite{sosso2016crystal},
and more importantly,
it is highly non-trivial to extract the values of the parameters that enter CNT 
using the microscopic quantities directly obtained from simulations.
For instance, the diffuse nature of solid-liquid interfaces makes it difficult to rationalize and formulate the nucleation free energy in a unique and meaningful way,
as the choice of the atomic order parameters used to distinguish ice structures affects the computed free energy profile
and the size of the critical nucleus~\cite{espinosa2016seeding,cheng2017gibbs,prestipino2018barrier}.
Furthermore,
an ice nucleus with complex stacking-disordered structures is usually formed during the homogeneous nucleation process ~\cite{malkin2012structure,haji2015direct,lupi2017role},
but it is computationally very expensive to exhaustively sample the numerous corresponding nucleation pathways even for nuclei that are within a small window of sizes~\cite{haji2015direct,lupi2017role}.
These multiple nucleation pathways make the theoretical analysis of the nucleation process even more difficult.

The present study establishes a rigorous and efficient framework to estimate the absolute homogeneous nucleation rate at realistic thermodynamic conditions,
and investigate how different physical quantities, including the interfacial free energy,
entropic contributions due to stacking disorder, and the kinetic prefactor contribute to the overall rate.
To demonstrate this framework on an important system with complex nucleation pathways,
we employ a monoatomic water (mW) model~\cite{molinero2008water},
which has proved very successful in reproducing many thermodynamic and structural properties of water including the melting point and the relative stability of different ice phases~\cite{hudait2016free} and as a result has been widely used to study ice nucleation~\cite{sosso2016crystal}.
We focus on a temperature $T=240~$K and pressure $P=1~$bar,
which are thermodynamic conditions commonly encountered in clouds~\cite{koop2000water}, freezers, and glaciers.

\subsection*{The nucleation free energy of ice Ih}

Rather than attempting to sample multiple nucleation pathways, we take an alternative route in which we first compute the nucleation rate for perfect ice Ih nuclei and then add correction terms to account for stacking disorder. 
This approach does not only allow us to fully converge the nucleation rate, but also disentangles contributions of different physical origins.
In order to restrict the sampling to the part of configuration space where only ice Ih nuclei form inside the liquid,
we opt for a combination of the umbrella sampling method and the seeding technique:
at the beginning of the simulation,
the fluid is seeded with an initial pure Ih ice cluster of a certain size.
At the same time, an umbrella potential is added to the Hamiltonian $\mathcal{H}(\textbf{q})$ of the system~\cite{torrie1977nonphysical},
so the biased Hamiltonian used in simulations is
\begin{equation}
    \mathcal{H}_{biased}(\textbf{q}) = \mathcal{H}(\textbf{q})
    + \dfrac{\kappa}{2} \left( \Phi - \bar{\Phi}\right)^2,
    \label{eq:H}
\end{equation}
where $\kappa$ denotes the spring constant of a rather stiff umbrella potential,
and the global collective variable (CV)
$\Phi=\sum_i \phi(i)$ is constructed by summing the
order parameter values $\phi(i)$ for each of the atoms in the system~\cite{cheng2015solid}.
Here, $\phi=S(\overline{Q_6})$ is the locally-averaged bond order parameter~\cite{lechner2008accurate}, which we transform with a hyperbolic switching function to enhance its resolving power between solid and liquid-like atomic environments.
The parameter $\bar{\Phi}$ is set to a value that ensures the stabilization of the initial ice Ih cluster at out-of-equilibrium conditions.
Repeating this procedure many times within a relevant range of sizes of the initial nuclei and different values of $\bar{\Phi}$,
one can reconstruct a free energy profile $\tilde{G}(\Phi)$ by using the WHAM method~\cite{souaille2001extension}.
\bc{Note that, as extensively discussed in Ref.~\citenum{cheng2016bridging}, 
using a global CV instead of the size of individual clusters to bias the simulation and construct the free energy profile is simpler and crucial for a rigorous determination of the nucleus size using a Gibbs dividing surface construction during the analysis.}

For the umbrella sampling simulations,
the NPT ensemble was employed throughout with the stochastic velocity rescaling thermostat~\cite{bussi2007canonical}
and an isotropic Nose-Hoover barostat.
A total of 8 sets of simulation runs with the biased Hamiltonian (Eqn.~\eqref{eq:H}) were performed using a system size of 8192 molecules.
\bc{For each set, about 50 umbrella sampling windows were used, and each trajectory lasted for about 0.5 ns.}
Fast implementation of this simulation setup was made possible by the flexibility of the PLUMED code~\cite{tribello2014plumed} in combination with LAMMPS~\cite{plim95jcp}.
The \sm contains annotated sample input files with detailed explanation of the simulation setups.

Note that the atomic order parameter $\overline{Q_6}$ does not distinguish ice Ic and ice Ih phases
but one can monitor the time evolution of another locally-average bond order parameter $\overline{Q_4}$ that is able to differentiate the two ice phases, which is plotted in the \sm.
One can also place a constraint on the number of particles that have Ic stacking.
That said, even without such explicit constraint very few stacking faults form during the umbrella sampling simulations, and almost all of them are in very small nuclei with fewer than 200 atoms.
\bc{The very low occurrence of the stacking faults here is probably due that the umbrella potential
places a strong constraint on the nucleus size in each simulation,
making it difficult for a basal bi-layer to dissolve and re-crystallize,
which is needed for the formation of a stacking disorder.}

\begin{figure}
\includegraphics[width=0.5\textwidth]{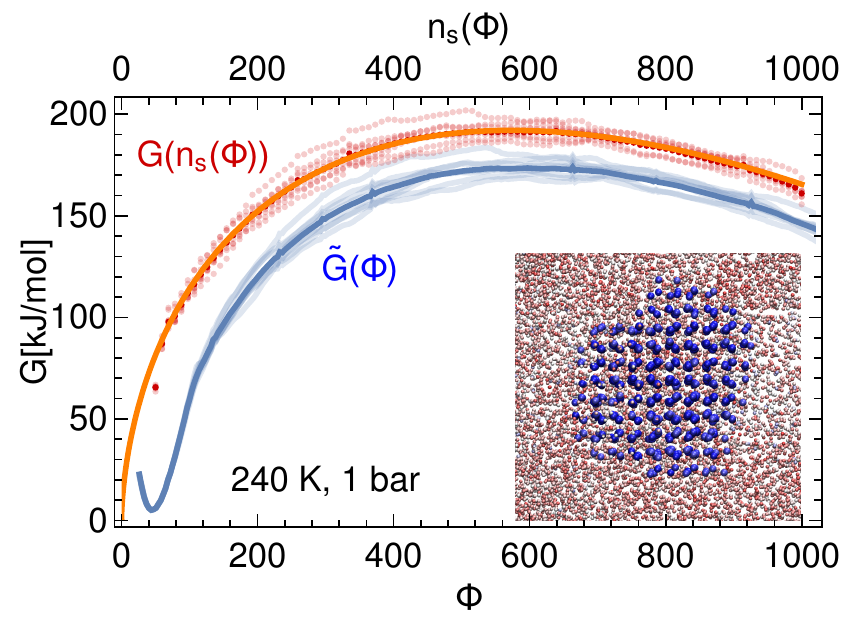}
\caption{
The light blue curves are the free energy profiles as a function of the collective variable
$\Phi$ for 8 sets of umbrella sampling simulations,
and the dark blue curve is the averaged $\tilde{G}(\Phi)$ from these runs.
Each dotted light red curve is the
free energy profile of a perfect Ih nucleus extracted from each set of simulations,
the dotted dark red curve is the averaged result,
and the orange curve indicates a CNT fit using Eqn.~\eqref{eq:cnttol}.
The inset shows a snapshot of an ice Ih nucleus embedded in liquid water.
}
\label{fig:gih}
\end{figure}

In Figure~\ref{fig:gih} we plotted the free energy $\tilde{G}(\Phi)$ that is associated with the formation of pure ice Ih nuclei.
From $\tilde{G}(\Phi)$,
one can extract the free energy profile $G_{\rm{Ih}}(\ns(\Phi))$ for the ice Ih nucleus as a function of the cluster size
by using the thermodynamic framework introduced in Refs.~\citenum{cheng2015solid,cheng2016bridging}.
Any extensive quantity\footnote{This extensive quantity might be the volume of the system, the energy,
or another global extensive collective variable~\cite{cheng2015solid,cheng2016bridging}.},
which is chosen to be the collective variable, $\Phi$ in this case,
can be used to unambiguously define a Gibbs dividing surface, which surrounds a nucleus and determines its size.
Reference~\citenum{cheng2016bridging} describes in detail 
the conversion between the free energy profile $\tilde{G}(\Phi)$ and 
the nucleation free energy as a function of
the size of the solid nucleus $\ns(\Phi)$.
We obtained $G_{\rm{Ih}}(\ns(\Phi))$ for the ice Ih nucleus with sizes larger than the cutoff value $n_{\rm{cut}}=50$
~\footnote{As discussed in Ref.~\citenum{cheng2016bridging}, the precise value of $n_{\rm{cut}}$ is not important, as it is a rough measure of what constitutes a solid nucleus rather than a local fluctuation.
In this case, we have also verified that choosing $n_{\rm{cut}}$ for values between 30 and 200 do not lead any noticeable change of the nucleation free energy.} 
and plotted the result in Figure~\ref{fig:gih}.

For a single reaction channel, 
the size $\ns(\Phi)$ alone is sufficient to characterize the nucleation free energy profile.
In such a case, 
CNT is commonly used to rationalize nucleation.
With a $\Phi$-based dividing surface,
the free energy of the ice Ih nucleus relative to the bulk liquid can be naturally decomposed into a bulk and a surface term~\cite{cheng2017gibbs},
\begin{equation}
    G_{\rm{Ih}}(\ns(\Phi))
    \eqsim \mu_{\rm{Ih}} \ns(\Phi) + \gp^{\Phi}\Omega \vs^{\frac{2}{3}}  \ns^{\frac{2}{3}}(\Phi)
    (1 +\zeta \ns^{-\frac{1}{3}}(\Phi)).
    \label{eq:cnttol}
\end{equation}
Here, $\mu_{\rm{Ih}}$ is the difference between the chemical potentials of ice Ih and the liquid phase,
$\gp^{\Phi}$ is the ice Ih-liquid interfacial free energy of the $\Phi$-based dividing surface at the planar limit,
and $\Omega$ is a geometrical constant.
The $\zeta \ns^{-\frac{1}{3}}(\Phi)$ term, which is determined by the distance between the $\Phi$-based dividing surface and the surface of tension in the planar limit,
captures the curvature dependence of the interfacial free energy $\gsl^{\Phi}$ of a curved interface.

The orange curve in Figure~\ref{fig:gih} shows a CNT fit to the nucleation free energy profile of the ice Ih nucleus using Eqn.~\eqref{eq:cnttol}.
The chemical potential $\mu_{\rm Ih}=-0.649$~kJ/mol at 240~K was obtained from previous calculations~\cite{espinosa2014homogeneous}, and
the other two parameters $\gp^{\Phi}=28.2(2)\, {\rm mJ/m^2}$ and $\zeta=0.2(1)$ were determined from the fit.
From this value of $\zeta$, we estimate the distance between the
$\Phi$-based dividing surface to the surface of tension ~\cite{cheng2017gibbs,cheng2018communication} to be 
$d=-\zeta(3\vs/32\pi)^{1/3}=0.2\, {\rm \AA} $.
For comparison, if all three parameters are used in the fit, the predicted values are $\mu_{\rm Ih}=-0.69(3)$~kJ/mol,
$\gp^{\Phi}=30(2)\, {\rm mJ/m^2}$ and $\zeta=0.0(1)$. 
\bc{As discussed in Ref.~\cite{cheng2017gibbs}, determining independently some of the parameters in the CNT expression reduces dramatically the uncertainties in the fit. }

The surface energy for the ice-liquid interface was computed to be around $35\, {\rm mJ/m^2}$ for the mW model at its melting point $\tm=274.6$~K~\cite{espinosa2016ice,ambler2017solid}.
In general, the interfacial free energy exhibits a temperature dependence,
and is also dependent on the specific choice of the extensive quantity used to define the Gibbs dividing surface away from the melting point~\cite{cheng2015solid}.
\bc{
However, as extensively discussed in Ref.~\citenum{cheng2017gibbs}, regardless
which dividing surface is used, as long as it is used consistently and the curvature correction term $\zeta$ is included
in the formulation of the nucleation free energy profile, no discrepancy will emerge.}
In this case, the correction term $\zeta$ is relatively small such that omitting it altogether from the CNT fit does not lead to significant changes in the estimation of the interfacial free energy,
however,
bear in mind that this may not be the case when using a different Gibbs dividing surface or when studying a different system.
In general, 
employing an interpolation or extrapolation using the naive version of the CNT without the curvature correction,
as is often the case in computational studies of this kind~\cite{espinosa2016seeding,li2011homogeneous},
may result in a systematic error in the prediction of the interfacial free energy as well as the nucleation barrier~\cite{cheng2017gibbs}.

\subsection*{Accounting for stacking disorder}
\begin{figure}[hbt]
\includegraphics[width=0.5\textwidth]{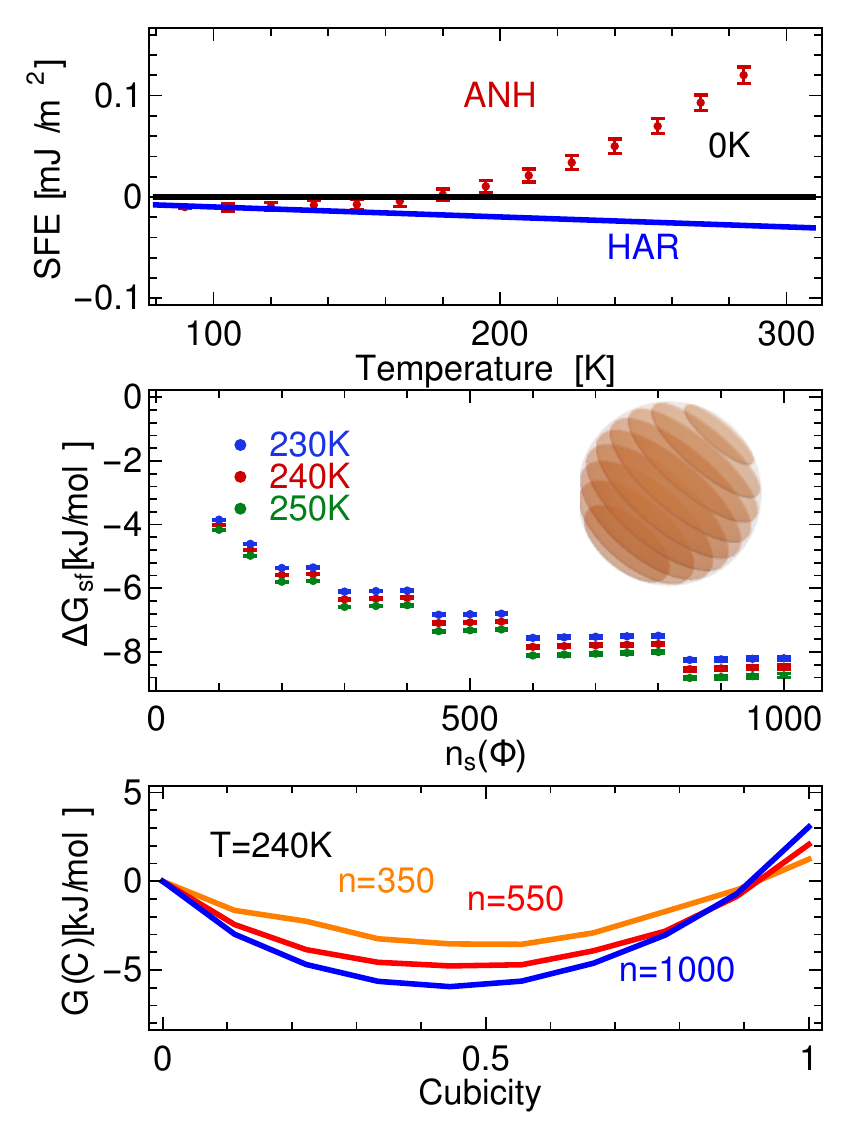}
\caption{
Upper panel: The stacking fault free energy per area as a function of temperature.
The black line is the estimate from the potential energy difference at 0 K, 
the blue curve represents the harmonic approximation (HAR), 
and the red dots show the results from thermodynamic integration that considers anharmonicity (ANH).
Statistical uncertainties are indicated by the error bars.
Middle panel: 
the free energy difference $\Delta G_{\rm sf}$ between a pure ice Ih nucleus and a one that has the same size and a mixed
stacking disorder at temperatures 230 K, 240 K and 250 K as predicted by the analytic model illustrated in the inset.
Lower panel:
The free energy profile as a function of cubicity for nuclei of three different sizes at 240 K.
}
\label{fig:sfe}
\end{figure}

In the previous section, we have considered a single nucleation pathway to form a pure ice Ih nucleus in liquid water.
We now formulate 
the difference in free energy of stacking-disordered and hexagonal
ice crystallites.
Because the ice-water interfacial free energies are indistinguishable for ice Ih and Ic phases ~\cite{ambler2017solid},
the free energy difference due to the formation of stacking disorders can only be associated with the bulk term.
Assuming the nucleus has a certain shape (e.g. spherical),
the number of bi-layers $k_{\rm max}$ along the basal plane can be easily calculated for a nucleus of
a given size (see the inset of Figure~\ref{fig:sfe}). 
\bc{The area of each plane $A_k$ at each  bi-layer can also be determined analytically as the bi-layers are equally spaced along the diameter of the spherical nucleus.}
Using an analytic model similar to the one-dimension Ising-like system discussed in Ref.~\citenum{quigley2014communication}:
We assume that the energy of each bi-layer only depends on its neighboring bi-layers,
thus on each layer the stacking order is independent,
and the probability $\rho_{k}$ of forming an ice Ic stacking order at the plane $k$ follows the Boltzmann distribution, i.e.,
\begin{equation}
    \rho_{k} = \dfrac{1}{Z_k}\exp \left(- \dfrac{A_k \gamma_{\rm sf}(T)}{ \kb T} \right),
    \label{eq:rhok}
\end{equation}
with the partition function for the $k$-th bi-layer
\begin{equation}
    Z_k = 1 + \exp \left(- \dfrac{A_k \gamma_{\rm sf}(T)}{ \kb T} \right),
    \label{eq:Z}
\end{equation}
where $\gamma_{\rm sf}(T)$ denotes the temperature dependent stacking fault free energy per unit area.
Note that the term stacking fault here refers to a stacking disorder with respect to the standard Ih stacking,
which means a pure Ic nucleus can be considered as having a stacking fault on every bi-layer.
The free energy difference between a pure ice Ih nucleus and one that has the same size and a stacking disordered structure can be expressed as
\begin{equation}
    \Delta G_{\rm sf} = -\kb T
    \ln  \prod_{k} Z_k.
    \label{eq:gsf}
\end{equation}

Using Eqn.~\eqref{eq:gsf},
the problem of computing the free energy difference between a pure Ih stacking nucleus and a mixed one has been reduced to
characterizing the stacking fault free energy $\gamma_{\rm sf}(T)$, which we determine next. To do that, we have carefully selected a combination of multiple thermodynamic integration routes,
in order to take into account vibrational entropy and anharmonicity, and make it possible to disentangle the different contributions to the stacking fault free energy.

The Gibbs free energy for a perfect bulk ice Ih structure and of a Ih bulk crystal with two stacking fault layers were computed separately 
using a sequence of thermodynamic integration routes, following the general strategy outlined in Ref.\citenum{cheng2018computing}.
We started by computing the Helmoltz free energy of a real system at a low temperature (90 K) by thermodynamic integration starting from a reference harmonic crystal.
Afterwards, we switched from
the NVT to the NPT ensemble and obtained the Gibbs free energy.
Finally, independent molecular dynamics simulations were performed in the NPT ensemble at temperatures ranging from 90 K to 300 K,
in order to obtain the temperature dependence of the Gibbs free energy for each system by thermodynamic integration with respect to $T$.

Figure~\ref{fig:sfe} shows that, considering only the potential energy difference at 0~K or under the harmonic approximation, the stacking fault free energy per unit area $\gamma_{\rm sf}(T)$ is estimated to be zero or negative. Only when anharmonicity is taken into consideration, $\gamma_{\rm sf}(T)$ is predicted to be positive, indicating that the ice Ih phase is more stable than Ic. 
This observation highlights the importance of accounting for anharmonic effects in studies of this kind.
In Reference~\citenum{hudait2016free}, $\gamma_{\rm sf}(\tm)$ was predicted to be $0.11(2)~\rm {mJ/m^2}$ by the method of growing ice bi-layers in simulations at near coexistence conditions,
which agrees well with our results using thermodynamic integration.
Notice also that the magnitude of $\gamma_{\rm sf}(T)$ is very small at all the temperatures considered here, which means that
for small ice nuclei the term in the exponent in the partition function $Z_k$ (Eqn.~\eqref{eq:Z}) is very close to zero.

Using the values of $\gamma_{\rm sf}(T)$, $\Delta G_{\rm sf}$ obtained from Eqn.~\eqref{eq:gsf} as a function of the size of the nucleus at three different temperatures is also shown in Figure~\ref{fig:sfe}.
The magnitude of the correction term $\Delta G_{\rm sf}$ is larger at high temperature because
it mostly stems from the entropic gain of forming stacking disorders,
but it is relatively insensitive with respect to temperature.
Because the analytic model introduced above accounts for the distribution of the stacking faults at each layer
using the accurate values of $\gamma_{\rm sf}(T)$,
one can enumerate all the possible combinations of the stacking disorder sequences,
and thereby compute the free energy as function of cubicity for a nucleus of a given size.
The cubicity here is defined as the fraction of ice molecules that are in local Ic environments.
The lower panel of Fig.~\ref{fig:sfe}
shows that for the three sizes considered here it is more favorable for the ice nucleus to adopt a cubicity close to $0.5$,
which is fully consistent with the observation in Ref.~\citenum{li2011homogeneous}. 
For the three nucleus sizes considered here, the free energy gain associated with the mixed stacking disorder is larger for the larger nuclei.
The most favorable degree of cubicity decreases with cluster size, consistent with the fact that for ice nuclei approaching macroscopic size,
the term $A_k \gamma_{\rm sf}(T)$ in Eqn.~\eqref{eq:Z} will dominate,
and pure Ih stacking should become favorable.
Note also that cubicity$=1$ corresponds to a ice Ic nucleus,
and thus the analytic model here predicts that the chemical potential difference $\mu_{Ic-Ih}$ between Ic and Ih is $0.0026(4)~{\rm kJ/mol}$ at 240~K.
This value agrees well with $\mu_{Ic-Ih}(240)=0.0031(2)~{\rm kJ/mol}$ that we computed independently using the thermodynamic integration method.
This agreement validates our assumption that
the each basal plane can be considered independently when calculating the free energy difference associated with stacking disorders.
Note that the analytic model here neglects the possibility of intersecting stacking disorders:
for ice Ih lattice stacking disorder can only occur along the two basal faces,
but in the uncommon cases when the nucleus consists a large enough domain of ice Ic, 
stacking can occur along the four (111) planes of the Ic lattice. 
It is also possible to construct a 2D Ising-like model to mimic the intersecting stacking disorders~\cite{lupi2017role},
although one has to make assumptions on the free energy penalty of a grain boundary-like structure
in the intersection.

\begin{figure}
\includegraphics[width=0.5\textwidth]{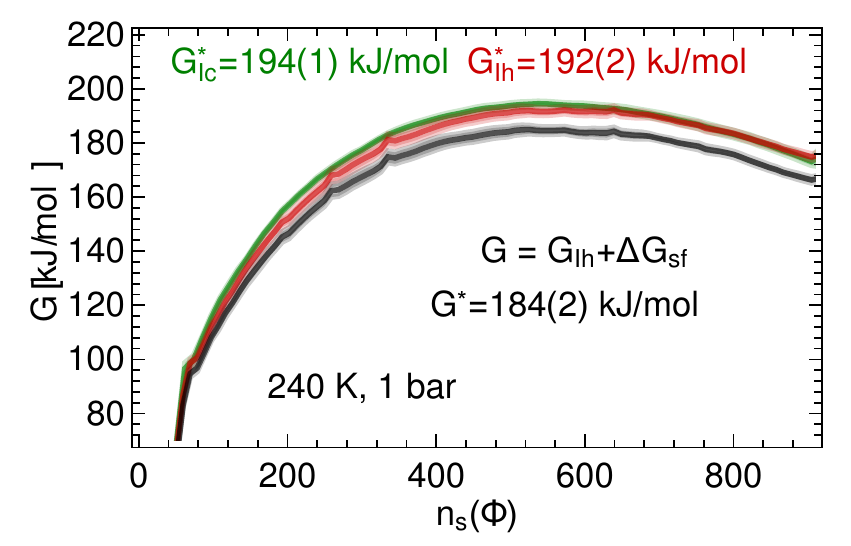}
\caption{
The red curve is the
free energy profile $G_{\rm{Ih}}$ of a pure Ih nucleus,
the green curve is $G_{\rm{Ic}}$ of a pure Ic nucleus,
and the black curve indicates the
free energy profile of an ice nucleus that can have a mixed stacking order.
The width of the curves indicates the statistical error in the free energy estimation,
computed from the error of the mean from independent simulation runs.
}
\label{fig:g}
\end{figure}

The cubicity$=1$ (i.e. pure ice Ic nucleus) case is illuminating also because one can perform a set of seeding and umbrella sampling calculations described in the previous section for the nucleation pathway of pure Ic nuclei. 
In Figure~\ref{fig:g} we report the free energy profile $G_{\rm{Ic}}$ as a function of the nucleus size.
The difference in the nucleation barriers $G^{\star}_{\rm{Ic}}-G^{\star}_{\rm{Ih}}$ between the Ic and the Ih nucleus is estimated to be $2 \pm 2~{\rm kJ/mol}$,
and the sizes of the critical nuclei (about 570 molecules) are similar in both cases.
This difference in nucleation barriers can be entirely explained by the chemical potential difference between the phases of ice $\mu_{Ic-Ih}(240\, {\rm K})$. 
This result thus supports our assumption as well as the conclusions in previous calculations~\cite{ambler2017solid} that the ice-water interfacial free energies are indistinguishable for ice Ih and Ic phases.
In Ref.~\citenum{lupi2017role} $G^{\star}_{\rm{Ic}}-G^{\star}_{\rm{Ih}}$ was estimated to be about $6 ~{\rm kJ/mol}$ at 230~K using the method of transition path sampling over ice nuclei whose sizes are close to that of the critical nucleus. 
The difference between their result and ours is still comparable to the magnitude of the statistical error,
and might be related to the fact that the Ic nuclei in our case were initially seeded in the undercooled liquid and therefore almost defect-free, but the ones in their study were generated by \bc{sampling dynamical trajectories} so may contain a higher concentration of defects.

Further comparisons with Ref.~\citenum{lupi2017role} can be made regarding the free energy profile as a function of cubicity (lower panel in of Figure~\ref{fig:sfe}).
Our analytic model, although simple, is able to capture the overall trend in the dependence of the free energy on cubicity. 
Our estimation for the free energy difference associated with staking disorders $\Delta G_{\rm{sf}}=8 ~{\rm kJ/mol}$ of the critical nucleus is smaller compared to the prediction $14~{\rm kJ/mol}$ in Ref.~\citenum{lupi2017role}.
The discrepancy may be due to that our analytic model neglects intersecting stacking disorders, grain boundaries, and domains of other phases of ice, which may further lower the free energy of ice nuclei.
In any case, this difference in $\Delta G_{\rm{sf}}$ here would only accounts for about one order of magnitude change in the estimated nucleation rate $J$.
It is also worth pointing out that it has been debated that the mW model underestimates the free energy
penalty associated with the Ic structures~\cite{quigley2014communication},
and one advantage of using an analytic model is the possibility to employ the experimental values or \emph{ab initio} results for $\gamma_{\rm sf}$ when estimating $\Delta G_{\rm{sf}}$ as well as the cubicity of nuclei of different sizes~\cite{quigley2014communication}.
We also want to point out that one has the option to sample the near critical ice nucleus with stacking disorder and other defects,
using transition path sampling ~\cite{li2011homogeneous,lupi2017role} or Monte Carlo methods~\cite{quigley2014communication},
in order to compute $\Delta G_{\rm{sf}}$.
Regardless how one chooses to evaluate $\Delta G_{\rm{sf}}$,
it can later be directly added on top of $G_{\rm{Ih}}$ for estimating the free energy profile of ice nucleus with defects, i.e.
$G(\ns(\Phi))=G_{\rm{Ih}}(\ns(\Phi))+\Delta G_{\rm{sf}}(\ns(\Phi))$.

\subsection*{The kinetic factor in homogeneous nucleation}

Once the free energy barrier of nucleation $G^\star=\max(G(\ns))$ has been determined,
the nucleation rate can be obtained from~\cite{auer2001prediction,espinosa2016time}
\begin{equation}
    J = (1/\vl) Z f^{+} \exp(-G^\star/k_BT) 
    \label{eq:J}
\end{equation}
where $\vl$ is the molar volume of the undercooled liquid, $f^+$ is the
addition rate of particles to the critical nucleus,
and the Zeldovich factor 
\begin{equation}
    Z = \sqrt{\frac{1}{2\pi k_B T}\dfrac{d^2 G(\ns)}{d n_s^2}}
\end{equation}
can be obtained numerically from the nucleation free energy profile $G(\ns(\Phi))$ in Figure~\ref{fig:g}. 
The addition rate $f^+$ can be computed as a diffusion coefficient from the mean square displacement of the cluster size after it is released at the top of the nucleation barrier~\cite{auer2001prediction,espinosa2016time}.
\bc{However, this approach assumes that $dG(\ns)/d\ns$ is effectively zero when running multiple trajectories,
is influenced by the choice of the initial configuration,
and by the latent heat created when the solid nucleus changes size~\cite{rozmanov2012isoconfigurational}.}
In order to overcome these shortcomings, we computed $f^+$ accurately and directly from the umbrella sampling trajectories with the biased Hamiltonian Eqn.~\eqref{eq:H} by applying a stochastic model 
originally proposed to mimic the kinetics of planar interfaces~\cite{pedersen2015computing}.

In the stochastic model, the time evolution of an extensive quantity $\Phi$ is viewed as resulting both from molecules changing from one phase to the other as well as from fluctuations in the bulk phases,
\begin{equation}
\label{eq:stochastic}
    \Phi(t) = (\phis-\phil) \ns(t) + \phil N + f(t).
\end{equation}
Here, $\phis$ and $\phil$ are the averages of the atomic order parameter in the bulk solid and the bulk liquid phases, respectively. The first term in the above equation stems from the temporal change of the size $\ns(t)$ of the solid cluster evolving under the action of the biased Hamiltonian. The term $f(t)$, on the other hand, takes into account fluctuations that do not change the composition of the solid-liquid system, but are due to changes of the extensive quantity $\Phi$ within the bulk phases, for instance caused by phonons propagating through the system. 

\begin{figure}
\includegraphics[width=0.5\textwidth]{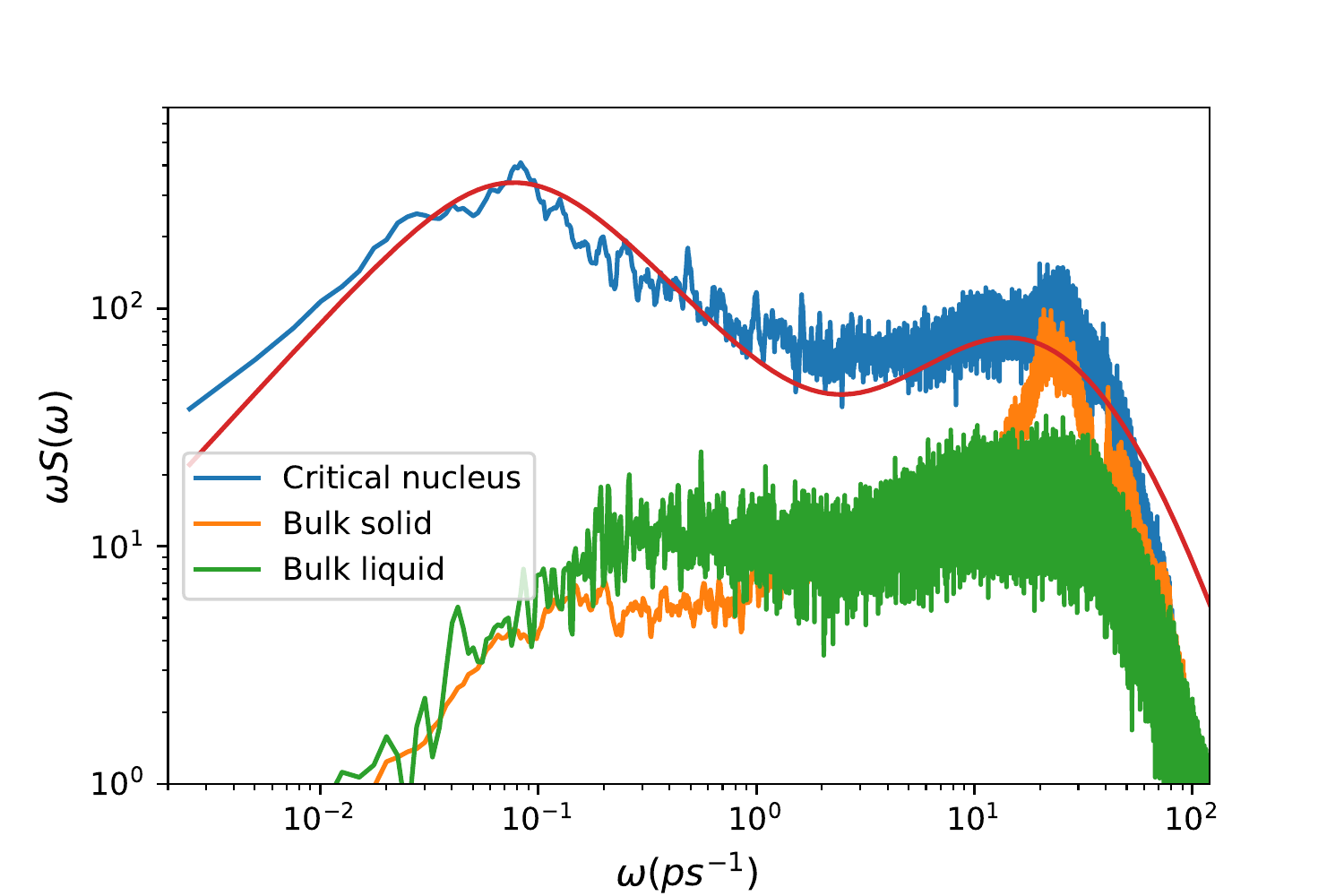}
\caption{
The orange, green and blue curves are the spectra $\omega S(\omega)$ for bulk solid, bulk liquid, and a solid-liquid system that contains a solid critical nucleus ($n^{\star}=550$) at 240~K and 1~bar, respectively.
The red curve is the fitting curve using Eqn.~\eqref{eq:somega} with parameters $m_f=3\times10^{-5}~{\rm ps^2 \,kJ/mol}$, $\kappa_f=0.03~{\rm kJ/mol}$, $\gamma_f=0.0025~{\rm ps \,kJ/mol}$,
and $\gamma=0.06~{\rm ps \, kJ/mol}$.
}
\label{fig:sww}
\end{figure}

In general, the time evolution of these two terms in Eqn.~\eqref{eq:stochastic} occurs on distinct time scales, as the change in $\ns(t)$ is determined by the relatively slow growth rate of the solid-liquid interface, and the dynamics of $f(t)$ happens on the time scale of lattice vibrations. These different time scales are reflected in the power spectrum $S(\omega)$ of $\Phi(t)$, related to the time autocorrelation function $\avg{\Phi(0) \Phi(t)}$ by 
\begin{equation}
    S(\omega) = \int_{-\infty}^\infty \avg{\Phi(0) \Phi(t)} e^{-i\omega t} dt.
\end{equation}
In Figure~\ref{fig:sww} we plot $\omega S(\omega)$ obtained for a solid-liquid system that contains a critical solid nucleus (blue  curve). For comparison, we also show the results for the bulk solid as well as bulk the liquid under the same thermodynamic conditions. It can be seen that for all three systems there is a peak at high frequency, which corresponds to the fast fluctuations. Only for the solid-liquid system with a critical nucleus there is another well separated peak at a low frequency, which stems from the growth of the crystalline nucleus embedded in the liquid.

To rationalize the power spectrum $S(\omega)$ further and extract quantitative information on the growth process from it, we now postulate that the time evolution of the collective variable $\Phi(t)$ can be modeled using a pair of coupled Langevin equations as described in Ref.~\citenum{pedersen2015computing}:
\begin{eqnarray}
    \gamma \dot{q} &=&
    - \kappa(f+q-\bar{\Phi})
    +\eta(t)\\
    m_f \ddot{f} &=&
    -\kappa_f f
    -\kappa(f+q-\bar{\Phi})
    -\gamma_f \dot{f}
    +\eta_f(t),
\end{eqnarray}
where the variable $q$, representing the slowly evolving part of $\Phi$, is defined as $q=(\phis-\phil) \ns(t) + \phil N$.
In the above equation, $\gamma$ and $\gamma_f$ are friction constants associated with $q$ and $f$, respectively,
$\eta(t)$ and $\eta_f(t)$ are Gaussian random forces. While the dynamics of $q(t)$ is assumed to be over-damped, inertial effects are included for the variable $f(t)$, which is assigned an effective mass of $m_f$. The force constant $\kappa$ is the sum of the umbrella spring constant of value $0.005\, {\rm kJ/mol}$ (Eqn.~\eqref{eq:H}) and the curvature of the free energy for the critical nucleus, $\kappa'=2d^2 G/d n_s^2=-0.0003\, {\rm kJ/mol}$, which is negligible in this case.
For this model, the power spectrum $S(\omega)$ of $\Phi(t)$ can be determined analytically,
\begin{multline}
        S(\omega) = \\
    \dfrac{2 k_B T}{\omega^2}{\rm Re}
    \left[\left[{\dfrac{1}{\gamma}+\left[{\gamma_f+i\left(\omega m_f-\dfrac{\kappa_f}{\omega}\right)}\right]^{-1}}\right]^{-1}-\dfrac{i\kappa}{\omega}\right]^{-1}.
    \label{eq:somega}
\end{multline}
We have fitted this expression to the power spectrum obtained from the umbrella sampling simulation and the result is shown in Figure~\ref{fig:sww}. As can be inferred from the figure, the simple Langevin model captures both peaks of the power spectrum, although the high-frequency peak is reproduced less accurately than the low-frequency peak, most likely because $f(t)$ is sensitive to the details of the order parameter in use.
However, we are primarily interested in the slow mode $q(t)$ and, particularly, in the value of the friction constant $\gamma$ because this parameter is related to the addition rate $f^+$ is by $f^+=k_B T / \gamma$.
Due to the separation of time scales, the parameters $m_f$ and $\kappa_f$ associated with the fast mode $f(t)$ have little bearing on the dynamics of the slow mode.
From the fit, we obtained $\gamma = 0.06~{\rm ps \,kJ/mol}$, yielding an 
addition rate of particles to the critical nucleus of $f^+ = 3 \times 10^{13}\,{\rm s^{-1}}$, which is smaller compared with previous results ($7 \times 10^{13}\,{\rm s^{-1}}$)~\cite{espinosa2016seeding} but of the same order of magnitude.
In addition, we estimated $f^+ = 2.5 \times 10^{13}\,{\rm s^{-1}}$ and $ 5 \times 10^{13}\,{\rm s^{-1}}$ for nuclei of about $340$ and $900$ atoms, respectively, confirming that the addition rate increases with nucleus size.

\section*{Results and discussions}

\begin{table*}
\caption{Key parameters for the estimation of the homogeneous ice nucleation rate at $T=240$~K and $P=1$~bar.
The value for the chemical potential $\mu_{\rm{Ih}}=-0.649 kJ/mol$ is from Ref.~\citenum{espinosa2014homogeneous}.}
\label{tab:gm}
\small
    \begin{tabular}{ c | c | c | c | c | c | c | c | c }
   \hline\hline
   type & $\msl$ [kJ/mol] & $\vs~$[\AA$^3$] & $\vl~$[\AA$^3$]  & 
   $G^\star$ [kJ/mol] & $Z$ & $f^+$ [s$^{-1}$] & $J$ [m$^{-3}$s$^{-1}$] & $\log_{10}$($J$m$^3$s)\\\hline   
 Ih &
 -0.649 & 30.48 & 29.79 &
 192(2) &
0.003 & 3$\times 10^{13}$ &
 $5\times10^{-3}$ & -2.3 \\
 
  Ic &
 -0.646 & 30.48 & 29.79 &
 194(1) &
0.004 & 3$\times 10^{13}$ &
 $2\times10^{-3}$ & -2.8 \\
 
   mixed &
 - & 30.48 & 29.79 &
 184(2) &
0.004 & 3$\times 10^{13}$ &
 0.3 & -0.5 \\
\hline \hline
    \end{tabular}
\end{table*}

Combining the free energy profile of stacking-disordered ice nucleus and the kinetic prefactor using Eqn.~\eqref{eq:J},
we estimate the nucleation rate to be $0.3~{\rm m^{-3}s^{-1}}$ at 240~K and 1~bar.
The key data for this estimation are tabulated in Table 1.
This estimation is about 9 orders of magnitude higher than the previous estimates using the seeding approach ($10^{-9} ~{\rm m^{-3}s^{-1}}$) ~\cite{espinosa2016seeding}.
A large part of the difference is probably due to a different definition for the size of the critical nucleus $n^\star$,
which is crucial in approximating the nucleation barrier if one employs the original expression of the classical nucleation theory for which $G^{\star}=n^\star \abs{\msl}/2$~\cite{espinosa2016seeding}.
It is common practice to set a threshold on the atomic order parameter in order to distinguish solid and liquid-like atoms and to determine the nucleus sizes~\cite{lupi2017role,espinosa2016seeding,espinosa2016time,prestipino2018barrier,li2011homogeneous,zimmermann2018nacl},
however, the nucleus size metric in use has a very strong influence on the estimated size~\cite{lupi2017role,prestipino2018barrier} and can thus affect the estimated rate by many orders of magnitude~\cite{zimmermann2018nacl}.
For example, with the $\Phi$-based Gibbs dividing surface we estimated $n^{\star}\approx 550$ molecules, and in Ref.~\citenum{espinosa2016time} the estimate is 688 molecules,
which implies that the estimated rate would differ by about 8 orders of magnitude using the original CNT expression.
It is worth pointing out that our framework eliminates this ambiguity in the determination of the critical nucleus and the nucleation rate,
making the classical nucleation theory much more rigorous.
Our estimated rate is also faster than the one estimated from forward flux sampling ($2 \times 10^{-7}~{\rm m^{-3}s^{-1}}$)~\cite{li2011homogeneous}.
Fortuitously, at 220~K,  the ice nucleation rate of mW water computed previously with umbrella sampling~\cite{reinhardt2012free,prestipino2018barrier} is also about 5 orders of magnitude higher than the corresponding forward flux sampling result~\cite{li2011homogeneous}.
\bc{In addition, for the case of a hard sphere system, the nucleation rates predicted by umbrella sampling and the ones predicted by forward flux sampling are more consistent though the former are still marginally faster at low supersaturations~\cite{filion2010crystal}.
Although the path-based techniques have a number of advantages including sampling real dynamics without the presence of bias potentials,
but there has been some worries that it may not proceed down the correct pathway that allows for sufficient local equilibration~\cite{reinhardt2012free}.}
For the mW model and for our choice of order parameter, not accounting for the influence from the curvature dependence of the interfacial free energy changes the predicted nucleation rate by 3 orders of magnitude in this case.
The effects from the curvature dependence, however, may be much larger for other systems or when using a different choice of the Gibbs dividing surface~\cite{cheng2017gibbs}.
Finally, we can estimate that the free energy associated with stacking disorders accelerates the nucleation rate by about 2 orders of magnitude. 

In most experimental measurements, the predicted homogeneous ice nucleation rate is around $10^{7} ~{\rm m^{-3}s^{-1}}$ ~\cite{taborek1985nucleation,demott1990freezing,pruppacher1995new}.
Our underestimation is probably due to the fact that the mW model overestimate the interfacial free energy,
i.e.,  $35~{\rm mJ/m^2}$~\cite{espinosa2016ice,ambler2017solid} compared with about $30~{\rm J/m^2}$ measured experimentally~\cite{granasy2002interfacial,hillig1998measurement} at $T_m$.
Note that, in general, atomistic studies of ice nucleation employing empirical water models suffer from various limitations.
As extensively discussed in Ref.~\citenum{espinosa2014homogeneous},
TIP4P models usually underestimate the chemical potential difference between the liquid and the ice phases when undercooled,
mW model overestimates the diffusion coefficient and some TIP4P models such as TIP4P/ice underestimates it, etc.

\section*{Conclusions}

In order to provide an accurate determination of the absolute nucleation rate for a monatomic water model, and to decompose it in physically-meaningful terms, we followed three key steps:
Firstly, we computed and characterized the free energy profile for pure ice Ih nuclei at an affordable computational cost.
Then we took into account multiple nucleation pathways due to the possibility of forming stacking disordered Ic layers in ice nuclei,
by adding an analytic free energy correction term.
We then calculated the kinetic prefactor using a stochastic model,
and finally obtained
the homogeneous ice nucleation rate using a general formulation of classical nucleation theory.

Our framework removed the ambiguity in defining the size of the nucleus,
and did not rely on many commonly-adopted approximations, such as neglecting the curvature dependence in interfacial free energy and the effect from stacking disorders,
both of which may have contributed to the long-standing controversy on the predicted nucleation rates, that varies by many orders of magnitude among previous studies.

The presented protocol for the study of homogeneous nucleation involving multiple nucleation pathway
\bc{makes fast and efficient computation of nucleation rates possible:
we performed a total of about 200 nanosecond simulations to obtain a well-converged free energy profile of nucleation with uncertainty estimations,
compared to tens or hundreds microseconds in the previous studies~\cite{lupi2017role,haji2015direct}.
}
More importantly,
disentangles all the terms that contribute to the overall rate,
including the difference in chemical potentials, the interfacial free energy, the stacking fault free energy,
and the kinetic prefactor.
Separating those terms not only greatly facilitates the theoretical understanding and formulation
of the homogeneous nucleation process,
but also enables one to evaluate these terms independently.
For instance, one can choose to compute the relatively important terms such as the chemical potential and interfacial free energy
employing an \emph{ab initio} potential energy surface~\cite{morawietz2016van,cheng2016nuclear}, perhaps even considering nuclear quantum effects which have been found to play in role in stabilizing the ice Ih phase~\cite{engel2015anharmonic},
and then combine them using the thermodynamic framework introduced here.
Our framework thus opens the door to the prediction of homogeneous ice nucleation rates using \emph{ab initio} potential energy surfaces, allowing for stringent cross-validations between theoretical and experimental estimates of this important quantity.
In addition, it also provides a recipe for how to tackle even more complex nucleation phenomena,
such as the crystallization of molecular crystals, the precipitation of halide salts, and the aggregation of hydrates.

\section{Supplementary Material}

The supplementary material contains a commented sample input file. 

\section{Acknowledgement}
We thank Federico Giberti
and David Wilkins
for critically reading the manuscript and making helpful suggestions.
We also thank Eduardo Sanz for helpful discussions and providing us with LAMMPS input files for mW water.
BC would like to acknowledge funding from the Swiss National Science 
Foundation (Project ID 200021-159896) and generous allocation of CPU time by
CSCS under Project ID s787. MC acknowledges funding by the European Research Council under the European Union's Horizon 2020 research and innovation programme (grant agreement no. 677013-HBMAP). CD acknowledges funding from the Austrian Science Fund (FWF) within the Spezialforschungsbereich Vienna Computational Materials Laboratory (Project F41). 

\end{document}